\documentclass[aps,prl,twocolumn,floatfix,bibliography,superscriptaddress]{revtex4}
\usepackage{graphics,bm,amsmath,amssymb,natbib,url,epsfig,psfrag,color}
\usepackage{graphicx}
\usepackage{subfigure}
\begin{document}
\newcommand{\Tr}{\text{Tr}}
\newcommand{\beq}{\begin{equation}}
\newcommand{\eeq}{\end{equation}}
\newcommand{\eran}[1]{{\color{blue}#1}}
\def\bs#1\es{\begin{split}#1\end{split}}	\def\bal#1\eal{\begin{align}#1\end{align}}
\newcommand{\nn}{\nonumber}
\newcommand{\sgn}{\text{sgn}}
\title{Measuring topological entanglement entropy using Maxwell relations}
	
	\author{Sarath Sankar}
\affiliation{School of Physics and Astronomy, Tel Aviv University, Tel Aviv 6997801, Israel}
	
	\author{Eran Sela}
	\affiliation{School of Physics and Astronomy, Tel Aviv University, Tel Aviv 6997801, Israel}
	
		\author{Cheolhee Han}
	\affiliation{School of Physics and Astronomy, Tel Aviv University, Tel Aviv 6997801, Israel}

\date{\today}
\begin{abstract}
	Topological entanglement entropy (TEE) is a key diagnostic of topological order, allowing to detect the presence of Abelian or non-Abelian anyons. However, there are currently no protocols to measure TEE	in condensed matter systems. Here, we propose a scheme to  measure the TEE, based on a nontrivial connection with the thermodynamic entropy change occurring in a quantum point contact (QPC) as it pinches off the topological liquid into two. We show how this entropy change can be extracted using Maxwell relations from charge detection of a nearby quantum dot. We demonstrate this explicitly for the Abelian Laughlin states, using an exact solution of the  sine-Gordon model describing the universal crossover in the QPC. Our approach might open a new thermodynamic detection scheme of topological states also with non-Abelian statistics.
\end{abstract}
\maketitle
\paragraph{Introduction.}
Entanglement describes non-local correlations between quantum objects and is  essential in understanding quantum many-body systems
.
It is also at the heart of quantum computation and information sciences and plays a pivotal role in models such as topological quantum computation~\cite{nayak2008non} and measurement based quantum computation~\cite{briegel2009measurement}.  Topological entanglement entropy (TEE)~\cite{kitaev2006topological,levin2006detecting,jiang2012identifying} is the ultimate 
diagnostic of 
topological order as defined by fractional quasiparticles carrying anyonic statistics. 
In a (2+1)-dimensional topological phase where the system size is larger than the correlation length, the entanglement entropy in the ground state $| \Psi \rangle$ between subsystems $A$ and $B$, $S_{EE} 
=-\Tr[\rho_A\log\rho_A]$ where $\rho_A=\Tr_B[|\Psi\rangle\langle \Psi|]$, takes the general form \cite{kitaev2006topological,levin2006detecting},
\begin{equation}
\label{eq:alphagamma}
	S_{EE}=\alpha L -\gamma+... ~.
\end{equation}
Here $L$ is the  length of the entanglement cut, $\alpha$ is a non-universal constant describing short range entanglement, and the second sub-leading term 
is  the TEE. The TEE $\gamma = \log \mathcal{D}$ is uniquely related to the total quantum dimension of the topological phase, $\mathcal{D}=\sqrt{\sum_a d_a^2}$, with $d_a$ being the  quantum dimensions of each individual anyon type labeled by $a$. In the presence of $N$ anyons of type $a$, the gapped topological liquid has a degeneracy that scales as $d_a^N$, so that $d_a>1$ ($d_a=1$) refers to a non-Abelian (Abelian) anyon. For instance, there are $m$ Abelian anyons in the fractional quantum Hall (FQH)  Laughlin state with filling fraction $\nu=1/m$, with $d_a=1$ $(a=0,1,\dots,m-1)$ and $\mathcal{D}=\sqrt{m}$; 
The Moore-Read state at $\nu=5/2$ has four abelian anyons and two non-Abelian anyons having $d_a=\sqrt{2}$, with $\mathcal{D}=2\sqrt{2}$.
	
Measuring entanglement entropy in many-body systems is a daunting task since it requires full state tomography. Variants of $S_{EE}$ such as the  R\'enyi entanglement entropy can be accessed in controllable many-body quantum simulators such as cold atoms or trapped ions using many-copy methods~\cite{daley2012measuring,pichler2013thermal,islam2015measuring,azses2020identification} or randomized measurement techniques~\cite{van2012measuring,elben2018renyi,brydges2019probing}. The latter, remarkably, was employed recently to measure the  R\'enyi TEE of Kitaev's toric code~\cite{kitaev2003fault} prepared on a quantum processor~\cite{satzinger2021realizing}. Unambiguously extracting the subleading TEE term $\gamma$ requires dividing the system into three subsystems and forming appropriate 
combinations of the different partitions such as to cancel the leading term~\cite{kitaev2006topological,levin2006detecting}. This was successfully implemented~\cite{satzinger2021realizing} thanks to the zero correlation length of the toric code
eigenstates. However, measuring TEE in condensed matter systems, such as in the realm of the FQH effect in two dimensional electron systems, remains elusive.

Interestingly, there exists an intimate relation between TEE and a thermodynamic entropy loss associated with a quantum point contact (QPC)~\cite{fendley2007topological}.  The QPC allows tunneling between two points on the edge (see parameter $\lambda$ in Fig.~\ref{fig:sidedQD} below). For an edge carrying fractional quasiparticles this is a relevant perturbation that introduces an energy scale $T_B$~\cite{kane1992transmission}. For $T \gg T_B$ the system is described by the ultraviolet (UV) fixed point that is unaffected by tunneling. As temperature decreases below $T_B$, tunneling processes proliferate and the system flows to the infrared (IR) fixed point where the droplet is disconnected into two droplets $A$ and $B$. Their anyonic charge $a$ can no longer fluctuate since each droplet can not support an overall fractional anyonic charge, leading to an entropy reduction. 
Using the bulk-edge correspondence, it was shown for a general (2+1)-dimensional chiral topological phase that the thermodynamic entropy change of the QPC coincides with a bulk property being precisely the TEE~\cite{fendley2007topological},
	\begin{equation}
		\label{eq:entropy_change}
		S_{UV}-S_{IR}=\log \mathcal{D}.
	\end{equation}
This entropy difference can be read out  from the temperature dependence of the entropy as shown in Fig.~\ref{fig:universalentropy} below. 
However, can this thermodynamic entropy change be measured experimentally?

Recently there has been a surge of interest in measuring entropy in mesoscopic systems
~\cite{yang2009thermopower,schmidt2017specific,hou2012ettingshausen,chickering2013thermoelectric,pyurbeeva2021controlling,kleeorin2019measure,pyurbeeva2021controlling,rozen2021entropic,saito2021isospin}.
Among those, a general entropy detection approach is based on charge measurements of quantum dots combined with Maxwell relations. While its promising applications were identified theoretically early on~\cite{cooper2009observable,yang2009thermopower,viola2012thermoelectric,ben2013detecting,ben2015detecting,smirnov2015majorana,sela2019detecting,hou2012ettingshausen,han2022fractional}, it was actually demonstrated very recently~\cite{kuntsevich2015strongly,hartman2018direct,child2022entropy,child2022robust}. 
   
In this paper, we rely on the general relation Eq.~(\ref{eq:entropy_change}) in the context of charge-based entropy detection experiments~\cite{hartman2018direct,child2022entropy,child2022robust}, to formulate a protocol to measure the TEE in FQH systems. 


\paragraph{Model.} 
The chiral edges of a FQH system described by an Abelian Laughlin state with filling factor $\nu=1/m$ 
are described by the bosonized Hamiltonian~\cite{wen1990chiral,SM}
\begin{equation}
	H_0=\frac{1}{4\pi\nu}\int
dx (\partial_x\varphi_L)^2+(\partial_x\varphi_R)^2,\label{eq:bareH}
\end{equation}
where $L/R$ denotes left/right movers. Here we have set the edge velocity to unity. The QPC at $x=0$ induces tunneling of quasiparticles between the edge states, which is described by~\cite{wen1991gapless,kane1992transmission} 
\begin{equation}
	H_B =\lambda\cos(\varphi_L(x=0)-\varphi_R(x=0)),\label{eq:backscattering} 
\end{equation}
where $\lambda$ is the tunneling strength. For $m>1$ it leads to the energy scale $T_B =C \lambda^{\frac{1}{1-\nu}}$ across which the crossover from UV to IR limits happen, with $C$ being a non-universal constant of appropriate dimensions~\cite{kane1992transmission}. The integer quantum Hall case $m=1$ is also described by the same Hamiltonian, but in this case the tunneling of electrons is marginal in the renormalization group sense and there is no crossover. While for experiments in electronic systems only odd values of $m$ are relevant, corresponding to fermions, we also consider theoretically the case of even $m$, corresponding to a Laughlin state of bosons.

\begin{figure}[t]
\includegraphics[width=\columnwidth]{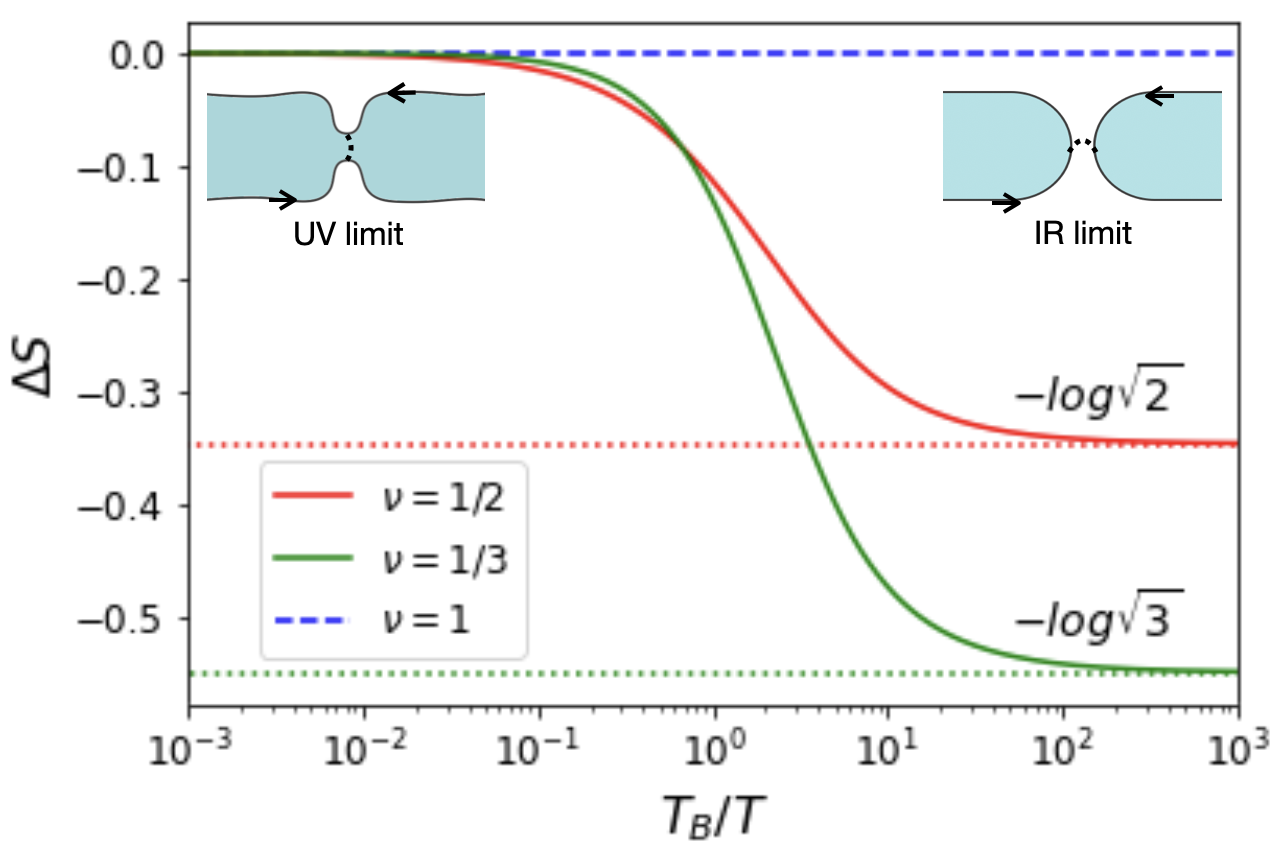}
\caption{\label{fig:universalentropy} Exact TBA results for the thermodynamic  entropy  change in a QPC at $\nu=1, 1/2, 1/3$ with a net entropy difference of $0$, $\log\sqrt{2}$, and $\log\sqrt{3}$, respectively. The results for $\nu=1/2$ are also obtained using refermionization~\cite{SM}.}
\end{figure}


In general for $\nu=1/m$, the model can be mapped into the boundary sine-Gordon model which is integrable~\cite{ghoshal1994boundary}. It can be solved using thermodynamic Bethe ansatz (TBA) ~\cite{fendley1994Exact,fendley1995exactprl,fendley1995Exact,SM}, which allows to obtain the free energy $F[T,T_B]$, and from it the boundary entropy $S=-\partial_T F$. 

Applying the TBA we compute the entropy along the full crossover from the UV to IR fixed points~\cite{SM}, see Fig.~\ref{fig:universalentropy}, which gives  $\mathcal{S}_{UV}-\mathcal{S}_{IR}= \log\sqrt{m}$. This yields a finite entropy change only in the fractional case, $m>1$. 
We included also the case $\nu=1/2$ which can be solved exactly using refermionization, and which corresponds to an effective Majorana fermion~\cite{guinea1985dynamics,kane1992transmission,matveev1995coulomb,smirnov2015majorana,SM}. Similar methods~\cite{schiller2020predicted,collaboratoin} can be applied to extract the entropy of parafermion modes.


\begin{figure}    
\centering    
\includegraphics[width= 0.74\linewidth]{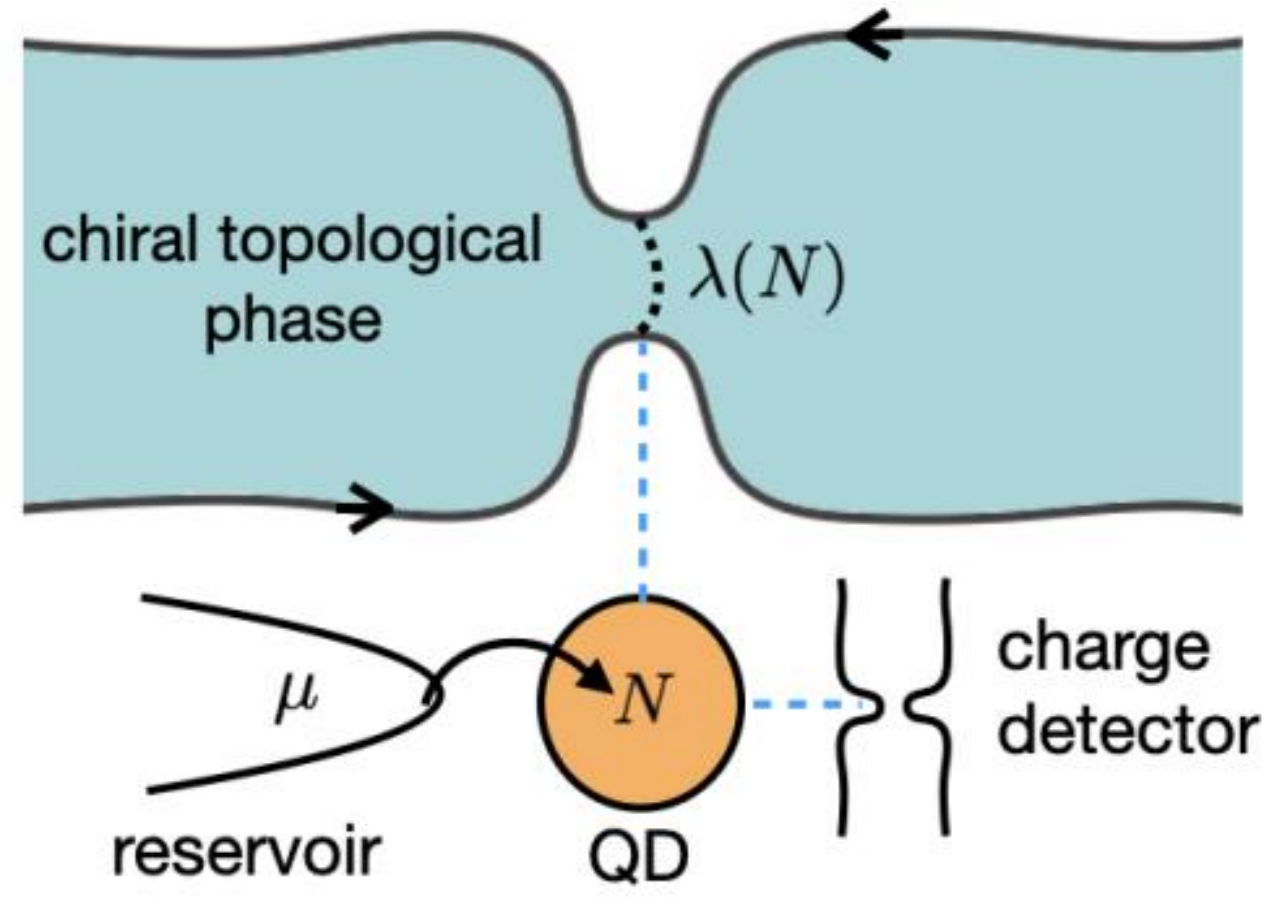}
\caption{
A QPC  in a chiral topological phase facilitates tunneling between the edge states. The tunneling strength $\lambda$ is controlled by the charge on a nearby QD, $\lambda=\lambda(N)$, that itself depends on the chemical potential $\mu$ of a nearby weakly tunnel-coupled reservoir. An adjacent charge detector measures the charge  of the QD $\langle N \rangle$. Sufficiently charging the QD drives the QPC between  UV to IR limits. }\label{fig:sidedQD}
\end{figure}

Next, we present two schemes to realize the crossover in the QPC between the UV and IR limits, and measure the resulting thermodynamic entropy change.

\paragraph{Scheme 1: Side coupled quantum dot.} We now illustrate the TEE measurement protocol 
for the $\nu=1/m$ Laughlin states using a side coupled quantum dot (QD) following the approach of Ref.~\onlinecite{sela2019detecting}.
As  shown in Fig.~\ref{fig:sidedQD}, we 
attach to the QPC a QD in the Coulomb blockade regime described by a classical energy function $E(N,\mu)=E_c N^2-\mu N$. Here $E_c$ is the charging energy,  $N$ is the number of electrons in the QD, and $\mu$ is a local chemical potential of the QD, controlled by a gate voltage. 
The QD interacts electrostatically with the QPC, as described by a dependence of the tunneling amplitude on the number operator of the QD, $\lambda = \lambda(N)$. Thus, the crossover energy scale $T_B$ is controlled by $N$, $T_B = T_B(N)$. We now show how, from such a dependence, one can extract $S_{UV}-S_{IR}$.

\begin{figure}
\centering
\includegraphics[width= 0.98\linewidth]{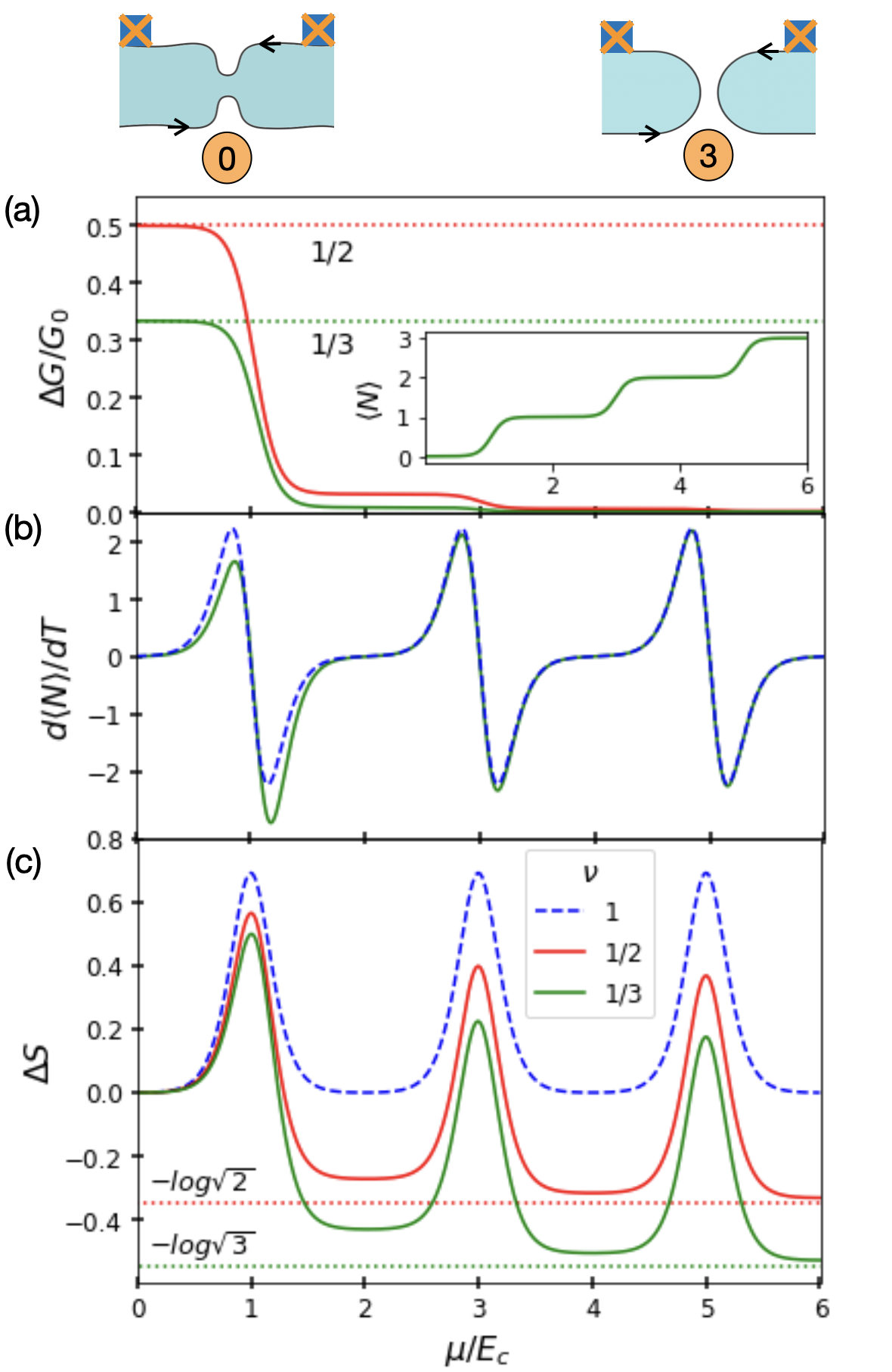}		
\caption{\label{fig:tee_measurement} Illustration of the proposed TEE measurement. (a) The conductance between the ohmic contacts for filling factor $\nu=1/2,1/3$ changes upon charging a nearby QD by varying the chemical potential $\mu$ of the QD. The charging curve is shown in the inset. Here $T/E_c=0.1$, $G_0=e^2/h$, and $T_B(N)/E_c=e^{N/2}-1$~\cite{sela2019detecting}. (b) Temperature derivative of the QD charge as a function of $\mu$. For $\nu=1$ (dashed) the curve is exactly anti-symmetric, while for $\nu=1/3$ (solid) it is not. (c) The resulting entropy change (QPC+QD), obtained by integrating $d \langle N \rangle /dT$ using the Maxwell relation Eq.~(\ref{eq:maxwell_relation}), for $\nu=1$ (dashed blue) and $\nu=1/2,1/3$ (solid curves). The peaks are associated with the $\log 2$ entropy of the dot at its charge steps. The plateaus of the entropy decrease for $\nu =1/2,1/3$ as a function of $\mu$, and converge to the desired TEE once the UV-IR crossover is accomplished.
}
\end{figure}

Under these conditions the  partition function of the combined system is
\begin{equation}
	\label{eq:tot_part_fun}
	Z_{\text{tot}}=\sum_N e^{-\frac{1}{T}[F(T,T_B(N))+E(N,\mu)]},
\end{equation}
where $F(T,T_B(N))$ is the free energy corresponding to the Hamiltonian in Eqs.~(\ref{eq:bareH}) and (\ref{eq:backscattering}) with $\lambda \to \lambda(N)$. 
By attaching a charge detector to the QD, using the Maxwell relation $\frac{d \langle N \rangle}{dT} = \frac{dS}{d\mu}$, one can extract the entropy change produced by a change of $\mu$,
\begin{equation}
	\label{eq:maxwell_relation}
	\Delta S_{\mu_1 \to \mu_2}=\int_{\mu_1}^{\mu_2}\frac{d\langle N \rangle}{dT} d\mu.
\end{equation}
When $T\ll E_c$, upon increasing $\mu$, there are several quantized charge steps in $N$
, see inset of Fig.~\ref{fig:tee_measurement}(a), and the QPC gets closer to pinch-off, corresponding to an increase in $T_B$. The desired entropy change will occur if, by charging the QD by $\Delta N$ electrons, the QPC transitions from the UV limit $(T_B \ll T)$ to the IR limit $(T_B \gg T)$.

 In Fig.~\ref{fig:tee_measurement}, we assume an $N$ dependence of $T_B(N)$ such that across a charging of $\Delta N=3$ electrons one achieves  $T_B(N) \ll T \ll T_B(N+\Delta N)$. In Fig.~\ref{fig:tee_measurement}(b) we compute $\frac{d \langle N \rangle}{dT}$, where $\langle N \rangle$ is extracted from the total free energy $F_{\text{tot}}=-T\ln Z_{\text{tot}}$ as $\langle N \rangle =-\partial F_{\text{tot}}/\partial \mu$. The total entropy change $S_{\mu_1=0 \to \mu}$ from a selected $\mu_1$ till a varying $\mu$ is shown in panel Fig.~\ref{fig:tee_measurement}(c). We can see that this entropy contains a series of $\log 2$ peaks corresponding to charge degeneracies of the QD, 
riding on top of the slowly varying entropy along the crossover for $\nu \ne 1$. For $T \ll E_c$ these two effects are well separated and 
we can measure the TEE by taking the difference between the corresponding entropy plateaus where $\langle N \rangle $ weakly fluctuates; see dashed horizontal lines in Fig.~\ref{fig:tee_measurement}(c).

Several experiments observed the crossover between UV and IR limits of a gate-tuned QPC for several FQH states through the conductance and the shot noise which crosses from quasiparticle tunneling to electron tunneling~\cite{chung2003anomalous,griffiths2000evolution,dolev2008observation,cohen2022universal}. 
Our method outlined in Fig.~\ref{fig:sidedQD}, however, requires to drive this crossover as a function of the chemical potential of the dot. The conductance for such a crossover is plotted in Fig.~\ref{fig:tee_measurement}(a). 

In practice, however, a side-coupled QD as in Fig.~\ref{fig:sidedQD} may have a limited electrostatic affect on the QPC. To enhance the effect, we now discuss a different setup with the QD embedded in the constriction.
	
\begin{figure}
	\centering
		\includegraphics[width=\linewidth]{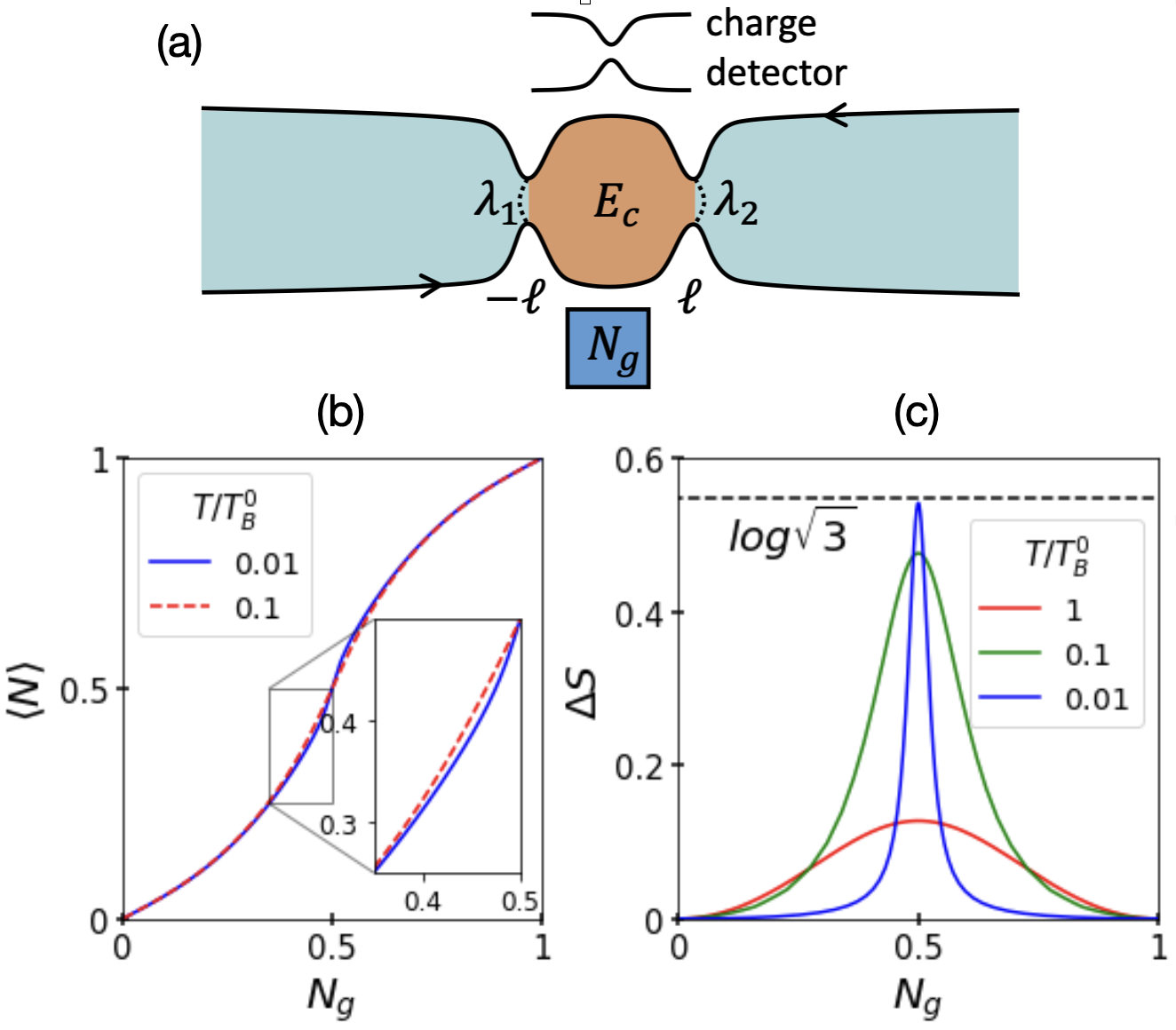}
		\caption{\label{fig:qd_setup} (a) Two QPCs in series creating a QD with Coulomb interaction $E_c$ within the FQH system. By tuning the gate voltage $N_g$ at fixed $\lambda_{1,2}$, this resonant double barrier-like system, yields an effective backscattering strength in Eq.~(\ref{eq:lambdaeff}) which drops at resonance and  vanishes for symmetric barriers. The charge detector measures the charge of the QD. (b) The average occupation of the QD for the case of symmetric barriers at $\nu=1/3$. Increasing temperature shifts the curve to right/left for $N_g$ above/below $1/2.$ (c) Entropy change inferred from Maxwell relation  for several ratios of $T/T_B^0$. As $T\to 0$, $N_g =0$ tends to the IR limit and the entropy difference at $N_g=1/2$ (which is always the UV limit)  tends to the TEE.}
	\end{figure}
	
\paragraph{Scheme 2: Coulomb blockade in the FQH regime.} 
One can realize a nearly complete change of transmission by replacing the QPC by a double barrier consisting of two QPCs, see Fig.~\ref{fig:qd_setup}(a). 
For non-interacting electrons it is well known~\cite{datta1997electronic} that the transmission has an abrupt resonance peak which is maximized for symmetric barriers. Here we demonstrate that this resonance effect also occurs in a double barrier system on a fractional edge. 
		
As shown in Fig.~\ref{fig:qd_setup}(a), we consider two QPCs in series separated by a distance $2\ell$, having backscattering amplitudes $\lambda_{1,2}$, and defining a 
QD characterized by a charging energy $E_c$. In this case the Hamiltonian is
\begin{align}
H_{QD}=&\lambda_1 \cos(\varphi_L(-\ell)-\varphi_R(-\ell))+\lambda_2 \cos(\varphi_L(\ell)-\varphi_R(\ell)) \nonumber\\
&+ \frac{E_c}{4\pi^2} \Big(\sum_{i=L,R}(\varphi_i(-\ell)-\varphi_i(\ell))-2\pi N_g\Big)^2.
\end{align}
In the limit of a large charging energy $E_c \ell \gg 1$ one obtains an effective $N_g$ dependent boundary condition between the bosonic field~\cite{lee2020fractional,morel2022fractionalization,lee2012visibility,SM}. As a result, the two QPCs behave  as one effective QPC with
\begin{equation}
\label{eq:lambdaeff}
H_B'=\lambda_{{\rm{eff}}}e^{i(\varphi_L(0)-\varphi_R(0))}+{\rm{h.c.}},
\end{equation}
where 
\beq
\label{eq:lambdaeffresult}
\lambda_{{\rm{eff}}}\propto \lambda_1 e^{-i \pi N_g}+\lambda_2 e^{i \pi N_g}.
\eeq
This effective model holds for $E_c\gg T_{B1,2}=C \lambda_{1,2}^{1/(1-\nu)}$.
We assume a finite reflection at the QPCs such that $T_{B1,2}  \gg T$. However, using Eq.~(\ref{eq:lambdaeffresult}), we see that when the barriers are symmetric $\lambda_1=\lambda_2\equiv\lambda_0$, we have $\lambda_{{\rm{eff}}} \propto  \lambda_0\cos(\pi N_g)$, or equivalently $T_B=T_B^0 | \cos(\pi N_g) |^{1/(1-\nu)}$. Hence  at $N_g=1/2$ the effective barrier vanishes and  $T_{B}=0$. 

Thus, the TEE can be extracted as the entropy reduction between the resonance condition $N_g=1/2$ and the off-resonance limit $N_g=0,1$. This entropy difference can be directly measured by attaching a charge detector to the  QD   and using the Maxwell relations. Eq.~(\ref{eq:maxwell_relation}) applies with $\mu \to 2 E_c N_g$.

In Fig.~\ref{fig:qd_setup}(b,c) we plot $\langle N \rangle$ and the extracted entropy. Here $\langle N \rangle =- \frac{1}{2E_c}\frac{\partial F}{\partial N_g}$ is computed from the TBA free energy $F(T,T_B)$ with $T_B=T_B[\lambda_{{\rm{eff}}}(N_g)]$ carrying the $N_g$-dependence via Eq.~(\ref{eq:lambdaeffresult}). Different than the side coupled QD, by using the resonance effect, the crossover is fully accomplished along the way from a Coulomb peak ($N_{g}=1/2$) and a nearby Coulomb valley ($N_{g}=0,1$). 
Also, different from the side coupled QD which was only weakly
coupled to the lead, and hence led to clear $\log 2$ entropy peaks at its charge degeneracy points [Fig.~\ref{fig:tee_measurement}(c)], here the strong tunnel-coupled QD has a negligible contribution to the entropy which scales as $S_{QD} \sim \frac{T_B^{2-2\nu}}{T^{2-2\nu}}\frac{ T^2}{E_c^2}$~\cite{SM}. However, in order to explore the full UV-IR crossover, the barriers should be high enough such that $T_B^0 \gg T$, see  Fig.~\ref{fig:qd_setup}(c).



\paragraph{Discussion.}
In the above schemes, to isolate the subleading TEE term $\gamma$ in Eq.~(\ref{eq:alphagamma}), we assume that when the FQH droplet  splits into two, the change in the length of the edge,  $L\to L+\Delta L$, leads only to a negligible entropy  change as compared to the order $\mathcal{O}(1)$ TEE.  We can now justify this assumption. 
	The resulting entropy change of the 1D gapless edge modes with velocity $v_F$  can  be estimated as
	\begin{equation}
	\label{deltaSL}
	\Delta S_L \sim \frac{ T }{\hbar v_F}\Delta L.
	\end{equation}
	Since the TEE part of the entropy change is $O(1)$, we can neglect $\Delta S_L$ if $\Delta L \ll \hbar v_F/T$. For $T=100 $ mK and $v_F=10^6 {\rm{m/s}}$, we find that the requirement $\Delta L \ll 10^{-4} {\rm{m}}$ is easily satisfied.

We also note that the TEE is a property of the ground state, which actually vanishes above a certain temperature determined by the energy gap to bulk excitations~\cite{lu2020detecting}. This restriction applies to our schemes to measure TEE as well.  

\paragraph{Summary.} Despite its importance, the measurement of the topological entanglement entropy is still elusive in condensed matter systems. 
Using a relation between TEE and a thermodynamic entropy change in fractional QPCs, we proposed realistic 
approaches to measure the TEE employing charge measurement and the Maxwell relation. We illustrated our protocols for the experimentally simplest and yet nontrivial case of Abelian fractional topological order.

Our proposed setups are also applicable to extract the entropy change between UV and IR fixed points in the more general boundary sine-Gordon model describing an impurity in a Luttinger liquid with any value of the interaction parameter, within mesoscopic systems simulating this model such as in Ref.~\onlinecite{anthore2018circuit}. 

Since the relation we applied  
is general, our method can be applied to more exotic  systems, such as complex hierarchical states~\cite{roosli2021fractional}, other topological systems such as spin liquid systems~\cite{savary2016quantum}, and most interestingly FQH states with non-Abelian quasiparticles.


	\emph{Acknowledgements.} We gratefully acknowledge support from the European Research Council (ERC) under the European Unions Horizon 2020 research and innovation programme under grant agreement
No. 951541. ARO (W911NF-20-1-0013) and the Israel Science Foundation grant numbers 154/19. We thank Anne Anthore, Christophe Mora, Yigal Meir, and particularly Frederic Pierre for illuminating discussions.

\end{document}


\newcommand{\beq}{\begin{equation}}
\newcommand{\eeq}{\end{equation}}
\newcommand{\bs}{\begin{split}}
\newcommand{\es}{\end{split}}
\def\bs#1\es{\begin{split}#1\end{split}}
\def\bal#1\eal{\begin{align}#1\end{align}}
\newcommand{\nn}{\nonumber}
\newcommand{\sgn}{\text{sgn}}
\pdfoutput=1

\title{Supplement: Measuring topological entanglement entropy using Maxwell relation}
	\author{Sarath Sankar}
\affiliation{School of Physics and Astronomy, Tel Aviv University, Tel Aviv 6997801, Israel}
	
	\author{Eran Sela}
	\affiliation{School of Physics and Astronomy, Tel Aviv University, Tel Aviv 6997801, Israel}
	
		\author{Cheolhee Han}
	\affiliation{School of Physics and Astronomy, Tel Aviv University, Tel Aviv 6997801, Israel}

\maketitle
\section{Mapping to the boundary sine-Gordon model and the boundary free energy from Thermodynamic Bethe Ansatz}
In this section, we explain the details of our bosonization convention and the mapping to the boundary sine-Gordon (BSG) model. From now on, we assume the velocity $v_F=1$, $\hbar=1$, and $k_B=1$.

We start by writing the Hamiltonian in Eqs.~(3) and (4) in the main text,
\beq\bs
H_0=&\frac{1}{4\pi\nu}\int_{-L/2}^{L/2} dx[(\partial_x\varphi_R)^2+(\partial_x\varphi_L)^2],\\
H_B =&\lambda\cos(\varphi_L(x=0)-\varphi_R(x=0)),
\es\eeq
with commutation relations of two counter-propagating fields $\varphi_R$ and $\varphi_L$,
\bal
[\varphi_R(x),\varphi_R(y)]=i\pi\nu\sgn(x-y),\quad [\varphi_L(x),\varphi_L(y)]=-i\pi\nu\sgn(x-y).
\eal
The total charge of the right and left modes is
\bal
Q_R=\int_{-L/2}^{L/2} dx\frac{1}{2\pi}\partial_x\varphi_R(x),\quad Q_L=-\int_{-L/2}^{L/2}\frac{1}{2\pi}\partial_x\varphi_L(x).
\eal
Here $L$ is the system size.
Since the quasiparticle has a fractional charge $\nu$, the quasiparticle operator satisfies a commutation rule
\bal
[Q_i,\psi^\dagger_j(x)]=\nu\delta_{ij} \psi^\dagger_j(x).
\eal
Hence we write the quasiparticle operators as
\bal
\psi_R(x)=:e^{i\varphi_R(x)}:,\quad \psi_L(x)=:e^{i\varphi_L(x)}:,
\eal
where $:\cdots :$ denotes normal ordering. 

To map this model into boundary sine-Gordon model, we define two purely right moving modes, denoted as even and odd,
\bal
\phi^e(x-t)=\frac{1}{\sqrt{2\nu}}(\phi_R(x,t)+\phi_L(-x,t)),\quad \phi^o(x-t)=\frac{1}{\sqrt{2}}(\phi_R(x,t)-\phi_L(-x,t)),\label{eq:chiralfields}
\eal
for $-L/2<x<L/2$. Then the Hamiltonian becomes
\bal
\label{eq:Hchiral}
H=\frac{1}{4\pi\nu}\int_{-L/2}^{L/2}dx (\partial_x\bar{\phi}^o)^2+\lambda\cos(\sqrt{2}\phi^o(0))+\frac{1}{4\pi}\int_{-L/2}^{L/2}dx (\partial_x\phi^e)^2.
\eal
We now discard the decoupled $\phi^e$ field. Next we fold the $\phi^o$ field on the half line of $x>0$ by 
\bal
\phi^o_R(x,t)=\phi^o(-x+t),\quad \phi^o_L=\phi^o(x+t),\quad x\in(0,L/2).
\eal
Then we can define a non-chiral field for $x\in(0,L/2)$,
\bal
\varphi^o=(\phi_R^o+\phi_L^o),\quad \Pi^o=\frac{1}{4\pi\nu}\partial_x(\phi_R^o-\phi_L^o).
\eal
Then the commutator is
\bal
[\varphi^o(x),\Pi^o(y)]=i\delta(x-y),
\eal
and the Hamiltonian becomes
\bal
H^o=\int_0^{L/2}dx \frac{1}{8\pi\nu}(\partial_x\varphi^o)^2+8\pi\nu (\Pi^o)^2+\lambda\cos(\frac{1}{\sqrt{2}}\varphi^o(0)).
\eal
This is the BSG model as discussed and solved exactly in Ref.~\cite{fendley1995Exact}. 

The BSG model is known to be integrable~\cite{ghoshal1994boundary} and the boundary free energy that we are interested in can be obtained by the thermodynamic Bethe Ansatz (TBA) method as elaborated in~\cite{fendley1994Exact}. The quasiparticle structure for $\nu=1/3$ consists of  kink ($+$), antikink($-$) and breather ($b$) states and the dispersion relation in terms of rapidity is obtained from the self-consistent solution of the set of  equations:
	\begin{equation}
		\label{eq:qp_energy}
		\epsilon_i(\theta)=\sum_{j}N_{ij}\int d\theta' K(\theta-\theta')  \log(1+e^{\epsilon_j(\theta')}), 
	\end{equation}
	where the subscripts $i,j$ denote the quasiparticle states, $K(\theta)=1/(\pi\cosh(2\theta))$ and the matrix $N$ is symmetric with the only finite elements being $N_{+ b}=N_{-b}=1$. Also note that $\epsilon_i$ is the quasiparticle energy scaled by temperature. The boundary free energy is~\cite{fendley1994Exact,fendley1996unified}:
	\begin{equation}
		\label{eq:tba_free_energy}
		F(T,T_B)=-\frac{T}{\pi}\int d\theta\frac{\log(1+e^{\epsilon_+(\theta)})}{\cosh(2(\theta-\log(T_B/T)))}+\frac{T}{2}\log3.
	\end{equation}
	By solving Eq.~\eqref{eq:qp_energy} numerically, plugging it into Eq.~\eqref{eq:tba_free_energy} and taking the temperature derivative, we obtain the dependence of boundary entropy on the ratio $T_B/T$ which is shown in Fig.~1 in the main text. The resulting free energy is then also applied to obtain the charging curves in our various entropy detection schemes from which entropy is obtained from Maxwell relations.

\section{Thermodynamic entropy for $\nu=1/2$}
The $\nu=1/2$ Laughlin state describes a FQH effect  of bosons. While not directly relevant for the mesoscopic experiments, it has various realizations using cold atom systems~\cite{sorensen2005fractional,palmer2006high,hafezi2007fractional,cooper2013reaching,lkacki2016quantum,cornfeld2015chiral,strinati2017laughlin,palm2022snapshot}, arrays of light-matter cavities~\cite{hayward2012fractional}, or only photonic states~\cite{kapit2014induced}. Experimental signatures have been observed in rotating traps of cold atoms~\cite{gemelke2010rotating}. Moreover an effective model for the interface between $\nu=1/3$ and $\nu=1$ corresponds to $\nu=1/2$ Laughlin state~\cite{cohen2022universal,chamon1997distinct}.

As a member of the Laughlin series, the $\nu=1/2$ Laughlin state has a TEE of $\log \sqrt{2}$, which can be extracted as a thermodynamic entropy change in a QPC. The associated entropy crossover can be obtained exactly from the TBA free energy, as shown in Fig.~1 in the main text. Interestingly, this case  can also be mapped to free fermions~\cite{guinea1985dynamics,kane1992transmission,matveev1995coulomb}, yielding an analytically exactly solvable model, and the resulting TEE can be associated with a free Majorana fermion. 

To see this we can refermionize the theory Eq.~(\ref{eq:Hchiral}) by defining $\psi(x,t)=\sqrt{\frac{D}{2\pi}}:e^{i\sqrt{2}\phi^o(x,t)}:f$, with $f^\dagger=f$ being a Majorana zero mode (MZM)~\cite{guinea1985dynamics,kane1992transmission,matveev1995coulomb} and $D$ is the bandwidth. The fermionic fields obey the equal time commutation relations: $\{\psi(x),\psi^\dagger(x')\}=\delta(x-x')$,  $\{f,f\}=2$, and $\{\psi(x),f\}=0$. The fermionized odd boson Hamiltonian becomes 
\begin{equation}
	H^o_{\nu=\frac{1}{2}}=-i\int_{-\frac{L}{2}}^{\frac{L}{2}} dx
\psi^\dagger(x)\partial_x \psi(x)
+\sqrt{\frac{\pi}{2D}} \lambda(\psi(0)-\psi^\dagger(0))f,
\end{equation}
and it describes a free fermion in the presence of an impurity which is a MZM at $x=0$. The contribution of the MZM to the density of states is~\cite{emery1992mapping,fabrizio1995anderson}, $\rho(E)=\frac{1}{2}(T_B/\pi)/(E^2+T_B^2)$ with $T_B=2\pi \lambda^2/D$.
Using the relation, $F=-T\int_{-\infty}^{\infty}dE\,\rho(E)\log(1+e^{-|E|/T})$, we obtain the required boundary free energy $F(T,T_B)$. The associated entropy is
\begin{equation}
	S=\frac{1}{\pi}\left(\frac{T_B}{T}\right)^3\int_{-\infty}^{\infty}dE \frac{\log(1+e^{-|E|/T})}{(E^2/T^2+T_B^2/T^2)^2}-\frac{1}{2}\log 2,
\end{equation}
where we used the convention that $S=0$ at $T_B=0$, which is the UV limit.  Clearly in the IR limit $T_B \gg T$, the first term tends to zero and we see that $S_{UV}-S_{IR}=\log \sqrt{2}$ as expected.

\section{Derivation of Eq.~(9) in the main text  for $\lambda_{{\rm{eff}}}$}
In this section, we derive Eq.~(9) in the main text from Eq.~(7) in the main text. Using the chiral fields in Eq.~\eqref{eq:chiralfields}, the Hamiltonian becomes
\beq\bs
H_0=&\int_{-L/2}^{L/2} dx\Big[\frac{1}{4\pi}\big(\partial_x\phi^e\big)^2+\frac{1}{4\pi\nu}\big(\partial_x\phi^o\big)^2\Big],\quad H_C=\frac{E_c\nu}{2\pi}\Big(\phi^e(\ell)-\phi^e(-\ell)-\sqrt{\frac{2}{\nu}}\pi N_g\Big)^2,\\
H_B=&\frac{\lambda_1}{2}e^{i\sqrt{\frac{\nu}{2}}(\phi^e(\ell)-\phi^e(-\ell))}e^{i\frac{1}{\sqrt{2}}(\phi^o(\ell)+\phi^o(-\ell))}+\frac{\lambda_2}{2}e^{i\sqrt{\frac{\nu}{2}}(\phi^e(-\ell)-\phi^e(\ell))}e^{\frac{i}{\sqrt{2}}(\phi^o(-\ell)+\phi^o(\ell))}+h.c..
\es\eeq
We can see that the tunneling term, which now derives from two points in space at $x=\pm \ell$, not only the odd field appears, as in Eq.~(\ref{eq:Hchiral}), but also the even field. 
When $E_c$ is very large, we can neglect the fluctuation of the $\phi^e$ field. After we integrate out the $\phi^e$ field and assuming $E_c\ell\gg 1$ and $T\ell\ll 1$ (in units where the velocity $v_F=1$ and $\hbar=1$), the average becomes
\bal
\langle e^{\pm i\sqrt{\frac{\nu}{2}}(\phi^e(\ell)-\phi^e(-\ell))}\rangle\simeq \left(\frac{2\nu E_c \gamma}{D\pi }\right)^{\frac{\nu}{2}} e^{\pm i \pi N_g}.\label{eq:chargefieldavg}
\eal
Here $\gamma=e^{\bf C}$ with Euler Gamma ${\bf C}$, and $D$ is a bandwidth. The full expression is written in  Eq.~\eqref{eq:expval}. 
Hence the model becomes
\bal
H_{eff}\simeq \frac{1}{4\pi\nu}\int_{-L/2}^{L/2} dx(\partial_x\phi^o)^2+ \left(\frac{2\nu E_c \gamma}{D\pi }\right)^{\frac{\nu}{2}} \Big[\frac{\lambda_1}{2}e^{i\pi N_g} e^{i\sqrt{2}\phi^o(0)}+\frac{\lambda_2}{2}e^{i\pi N_g}e^{-i\sqrt{2}\phi^o(0)}+h.c.\Big].
\eal
Hence 
\bal
\lambda_{{\rm{eff}}}=\left(\frac{2\nu E_c \gamma}{D\pi }\right)^{\frac{\nu}{2}} \frac{1}{2}(\lambda_1 e^{i\pi N_g}+\lambda_2 e^{-i\pi N_g}).
\eal

\section{Derivation of Eq.~(\ref{eq:chargefieldavg})}
In this section, we derive Eq.~\eqref{eq:chargefieldavg}. The partition function of the $\phi^e$ fields without tunneling is
\bal
Z_{\phi^e}=\int \mathcal{D}\phi^e\exp\Bigg[\sum_{\omega_n,q}&-\frac{q^2-i\omega_n q}{4\pi}\phi^e(q,\omega_n)\phi^e(-q,-\omega_n)-\frac{E_c\nu}{2\pi^2}\sum_{n}(\phi^e(\ell,\omega_n)-\phi^e(-\ell,\omega_n))(\phi^e(\ell,-\omega_n)-\phi^e(-\ell,-\omega_n))\nn\\
&+\frac{E_c \sqrt{2\nu}N_g}{\pi\sqrt{T}}(\phi^e(\ell,0)-\phi^e(-\ell,0))-\frac{E_c N_g^2}{T}\Bigg].
\eal
Here we use the Fourier transform of $\phi^e(x,\tau)$ as
\bal
\phi^e(x,\tau)=\sqrt{\frac{T}{L}}\sum_{\omega_n,q}\phi^e(q,\omega_n)e^{iqx-i\omega_n\tau},\quad \phi^e(\tau)=\sqrt{T}\sum_{\omega_n}\phi^e(\omega_n)e^{-i\omega_n\tau},
\eal
with $\omega_n=2\pi T n$.
Our aim is to integrate out $\phi^e$ except at the points $x=\pm \ell$. We proceed by writing
\bal
Z_{\phi^e}=&\int \mathcal{D}\phi^e_\pm\int \mathcal{D}\phi^e e^{-S(\phi^e)}\delta(\phi^e(\ell,\tau)-\phi^e_+(\tau))\delta(\phi^e(-\ell,\tau)-\phi^e_-(\tau))\nn\\
=&\int \mathcal{D}\phi^e_\pm\int \mathcal{D}\lambda_\pm \int\mathcal{D}\phi^e e^{-S(\phi^e)-\int_0^\beta d\tau [\lambda_+(\tau)(\phi^e(\ell,\tau)-\phi^e_+(\tau))+\lambda_-(\tau)(\phi^e(-\ell,\tau)-\phi^e_-(-\tau))]}.
\eal
Here we expressed the $\delta$ functions using Lagrange multipliers $\lambda_\pm$.
Next we collect the $\phi^e$ dependent terms in the partition function, 
\bal
\int\mathcal{D}\phi^e \exp\Bigg[\sum_{\omega_n,q}&-\frac{q^2-i\omega_n q}{4\pi}\phi^e(q,\omega_n)\phi^e(-q,-\omega_n)-\frac{1}{2\sqrt{L}}\Big(\lambda_+(\omega_n)\phi^e(-q,-\omega_n)e^{-iq\ell}+\lambda_+(-\omega_n)\phi^e(q,\omega_n)e^{iq\ell}\Big)\nn\\
&-\frac{1}{2\sqrt{L}}\Big(\lambda_-(\omega_n)\phi^e(-q,-\omega_n)e^{iq\ell}+\lambda_-(-\omega_n)\phi^e(q,\omega_n)e^{-iq\ell}\Big)\Bigg].
\eal
After integrating out $\phi^e(q,\omega_n)$, the $\lambda_\pm$ dependent parts become
\bal
\int\mathcal{D}\lambda_\pm \exp\Bigg[\sum_{\omega_n,q}&\frac{4\pi}{q^2-i\omega_n q}\frac{1}{4L}(\lambda_+(\omega_n)e^{-iq\ell}+\lambda_-(\omega_n)e^{iq\ell})(\lambda_+(-\omega_n)e^{iq\ell}+\lambda_-(-\omega_n)e^{-iq\ell})\nn\\
&+\frac{1}{2}(\phi^e_+(\omega_n)\lambda_+(-\omega_n)+\phi^e_+(-\omega_n)\lambda_+(\omega_n)+\phi^e_{-}(\omega_n)\lambda_-(-\omega_n)+\phi^e_-(-\omega_n)\lambda_-(\omega_n))\Bigg].
\eal
First we focus on $\omega_n=0$. The equation becomes
\bal
\exp\Bigg[\sum_{q}&\frac{\pi}{q^2}\frac{1}{L}\Big(\lambda_+(0)\lambda_+(0)+\lambda_-(0)\lambda_-(0)+2\lambda_+(0)\lambda_-(0)\cos(2q\ell)\Big)+(\phi^e_{+}(0)\lambda_+(0)+\phi^e_{-}(0)\lambda_-(0))\Bigg].
\eal
It can be evaluated using the summation relations valid for $L\rightarrow\infty$,
\bal
\sum_q \frac{1}{q^2}&\simeq \frac{L^2}{12},\quad 
\sum_q \frac{e^{\pm i qx}}{q^2} \simeq\frac{L^2}{12}(1-\frac{6 x}{L}).
\eal
After integrating out $\lambda_\pm(\omega_n=0)$, we have
\bal
(\omega_n=0\text{ part})=\exp\Big[&\frac{1}{8\pi \ell}(\phi^e_{+}(0)-\phi^e_{-}(0))^2\Big].
\eal

Next we focus on $\omega_n\ne 0$. After integrating out $q$,
\bal
=\exp\Bigg[\sum_{\omega_n,q}&\frac{\pi}{2|\omega_n|}\big(\lambda_+(\omega_n)\lambda_+(-\omega_n)+\lambda_-(\omega_n)\lambda_-(-\omega_n)\big)-\Big(\frac{\pi}{2\omega_n}+\Theta(-\omega_n)\frac{\pi}{|\omega_n|}e^{\omega_n 2\ell}\Big)\lambda_+(\omega_n)\lambda_-(-\omega_n)\nn\\
&+\Big(\frac{\pi}{2\omega_n}-\Theta(\omega_n)\frac{\pi}{|\omega_n|}e^{-\omega_n 2\ell}\Big)\lambda_+(-\omega_n)\lambda_-(\omega_n)\nn\\
&+\frac{1}{2}(\phi^e_{+}(\omega_n)\lambda_+(-\omega_n)+\phi^e_{+}(-\omega_n)\lambda_+(\omega_n)+\phi^e_{-}(\omega_n)\lambda_-(-\omega_n)+\phi^e_{-}(-\omega_n)\lambda_-(\omega_n))\Bigg].
\eal
Here we used the integral
\bal
&\int_{-\infty}^{\infty} dq \frac{ \cos(q x)}{q(q-i\omega_n)}=\frac{\pi e^{-|\omega_n|x}}{|\omega_n|}.
\eal
After integrating out $\lambda_\pm (\omega_n)$, we have
\bal
(\omega_n\ne 0\text{ part})=&\exp\Bigg[\sum_{i\omega_n}-\frac{|\omega_n|}{4\pi(1-e^{-|\omega_n|2\ell})}\Big[\phi^e_{+}(\omega_n)\phi^e_{+}(-\omega_n)+\phi^e_{-}(\omega_n)\phi^e_{-}(-\omega_n)\nn\\
&+\Big(\frac{|\omega_n|}{\omega_n}+2\Theta(-\omega_n)e^{\omega_n 2\ell}\Big)\phi^e_{+}(\omega_n)\phi^e_{-}(-\omega_n)-\Big(\frac{|\omega_n|}{\omega_n}-2\Theta(\omega_n)e^{-\omega_n 2\ell}\Big)\phi^e_{+}(-\omega_n)\phi^e_{-}(\omega_n)\Big]\Bigg].
\eal
For convenience, we change the basis as $\phi_1^e=(\phi^e_+ +\phi^e_-)/\sqrt{2}$, $\phi_2^e=(\phi^e_+-\phi^e_-)/\sqrt{2}$, and integrate out $\phi_1^e$ field since it is decoupled with the Coulomb interaction. The partition function is
\bal
Z_{\phi^e}=\int \mathcal{D}\phi_2^e\exp\Bigg[\sum_{\omega_n}-\Big[\frac{|\omega_n|}{2\pi(1+e^{-|\omega_n|2\ell})}+\frac{E_c\nu}{\pi^2}\Big]\phi^e_2(\omega_n)\phi_2^e(-\omega_n)+\frac{\phi_2^e(0)^2}{4\pi \ell}+\frac{2E_c\sqrt{\nu}N_g}{\pi\sqrt{T}}\phi_2^e(0)-\frac{E_c N_g^2}{T}\Bigg].
\eal
Consider the expectation value,
\bal
\langle e^{\pm i\sqrt{\frac{\nu}{2}}(\phi^e(\ell)-\phi^e(-\ell))}\rangle=\int \mathcal{D}\phi_2^e\exp\Bigg[&\sum_{\omega_n}-\Big[\frac{|\omega_n|}{2\pi(1+e^{-|\omega_n|2\ell})}+\frac{E_C\nu}{\pi^2}\Big]\phi^e_2(\omega_n)\phi_2^e(-\omega_n)\pm i\sqrt{T \nu}\phi^e_2(\omega_n)\nn\\
&+\frac{\phi_2^e(0)^2}{4\pi \ell}+\frac{2E_c\sqrt{\nu}N_g}{\pi\sqrt{T}}\phi_2^e(0)-\frac{E_c N_g^2}{T}\Bigg].
\eal
After integrating out $\phi_2^e$, we obtain
\bal
\langle e^{\pm i\sqrt{\frac{\nu}{2}}(\phi^e(-\ell)-\phi^e(\ell))}\rangle=&\exp\Bigg[\frac{\pm i\pi N_g}{1+\frac{\pi}{4\ell E_c\nu}}
 \Bigg]\exp\Bigg[-\frac{\pi^2 T}{4E_c}\frac{1}{1+\frac{\pi}{4\ell E_c\nu}}-\sum_{\omega_n}\frac{-\pi T \nu/2}{\frac{|\omega_n|}{1+e^{-|\omega_n|2\ell}}+2E_c\nu/\pi}\Bigg].\label{eq:expval}
\eal
This expression is exact. Our regime of interest is 
\beq
\label{eq:regimeinterest}
E_c \gg \frac{1}{\ell} \gg T,
\eeq
i.e. the temperature is smaller than the level spacing within the QD, which is smaller than the charging energy. Then the first factor becomes $e^{\pm i \pi N_g}$. Unfortunately there is no closed expression for the summation, which also depends on the high energy cutoff $D$. However it can be approximately calculated. First we separate the summation into two parts as $\sum_{\omega_n}(\cdot)=\sum_{|\omega_n|>1/\ell} (\cdot)+ \sum_{|\omega_n|<1/\ell}(\cdot)$. 
In the first term with $|\omega_n|>1/\ell$, the factor $e^{-|\omega_n|2\ell}$ becomes negligibly small, hence the summand becomes $(\cdot) \to \frac{-\pi T \nu/2}{|\omega_n|+2E_c\nu/\pi}$.
On the other hand, in the second term with $|\omega_n|<1/\ell$, $|\omega_n|$ is much smaller than $E_c$ (since $|\omega_n|<1/\ell\ll E_c$), therefore 
the factor of $|\omega_n|$ is negligible compared to $E_c$ and we therefore approximate the summand by the same expression $(\cdot) \to \frac{-\pi T \nu/2}{|\omega_n|+2E_c\nu/\pi}$. 
The resulting summation in Eq.~\eqref{eq:expval} can the be evaluated, yielding
\bal
\langle e^{\pm i\sqrt{\frac{\nu}{2}}(\phi^e(-\ell)-\phi^e(\ell))}\rangle\simeq e^{\pm i\pi N_g}\exp\Bigg[-\frac{\pi^2 T}{4E_c}-\sum_{\omega_n}\frac{-\pi T \nu/2}{|\omega_n|+2E_c\nu/\pi}\Bigg]\simeq \left(\frac{2\nu E_c \gamma}{D\pi }\right)^{\frac{\nu}{2}}e^{\pm i\pi N_g}.\label{eq:expval1}
\eal
Here we neglect $\mathcal{O}(T/E_c)^2$ order. The high energy cutoff is implemented by adding a factor $e^{-|\omega_n|/D}$.

\section{Entropy correction}
 For a weakly coupled QD, at the charge degeneracy point $N_g=1/2$ there is an extra $\log 2$ entropy contribution. While it is suppressed for the strongly coupled dot, as discussed in scheme 2 in the main text, yet there is a finite entropy contribution which is not accounted by the TBA result.  In this section, we discuss this deviation of the entropy of QD system from TBA result with $\lambda \to \lambda_{{\rm{eff}}}$.

For this purpose, we compute the backscattering correction of the partition function near $N_g=1/2$, and estimate the entropy correction at finite temperature. Since 1st order correction vanishes, consider the 2nd order correction of the partition function. Assuming $\lambda_1=\lambda_2=\lambda_0$ for simplicity, we have
\bal
\frac{Z_2}{Z_0} =\frac{\lambda_0^2}{2} \sum_{\zeta_{1,2}=\pm1}\int_0^{1/T} d\tau_1d\tau_2 &\langle e^{i\sqrt{2}\phi^o(0,\tau_1)}e^{-i\sqrt{2}\phi^o(0,\tau_2)} \rangle \langle e^{i\zeta_1\sqrt{\frac{\nu}{2}}(\phi^e(-\ell,\tau_1)-\phi^e(\ell,\tau_1))}e^{i\zeta_2\sqrt{\frac{\nu}{2}}(\phi^e(-\ell,\tau_2)-\phi^e(\ell,\tau_2))}\rangle.
\eal
The $\phi^o$ factor gives
\bal
\langle e^{i\sqrt{2}\phi^o(0,\tau_1)}e^{-i\sqrt{2}\phi^o(0,\tau_2)} \rangle=\frac{(\pi T/D)^{2\nu}}{\sin^{2\nu}(\pi T(\tau_1-\tau_2))}. 
\eal
The contribution from the sector of $\phi^e$ is more complicated, and given by
\bal
&\langle e^{i\zeta_1\sqrt{\frac{\nu}{2}}(\phi^e(-\ell,\tau_1)-\phi^e(\ell,\tau_1))}e^{i\zeta_2\sqrt{\frac{\nu}{2}}(\phi^e(-\ell,\tau_2)-\phi^e(\ell,\tau_2))}\rangle\nn\\
=&\int \mathcal{D}\phi_2^e\exp\Bigg[\sum_{\omega_n}-\Big[\frac{|\omega_n|}{2\pi(1+e^{-|\omega_n|2\ell})}+\frac{E_c\nu}{\pi^2}\Big]\phi^e_2(\omega_n)\phi_2^e(-\omega_n)+ i\sqrt{T \nu}(\zeta_1 e^{i\omega_n\tau_1}+\zeta_2 e^{i\omega_n\tau_2})\phi^e_2(\omega_n)\nn\\
&+\frac{\phi_2^e(0)^2}{4\pi \ell}+\frac{2E_c\sqrt{\nu}N_g}{\pi\sqrt{T}}\phi_2^e(0)-\frac{E_C N_g^2}{T}\Bigg]\nn\\
=&\exp\Bigg[\sum_{\omega_n}\frac{-\pi(1+e^{-|\omega_n|2\ell})T\nu/2}{|\omega_n|+2E_c\nu/\pi (1+e^{-|\omega_n|2\ell})}(2+2\zeta_1\zeta_2\cos(\omega_n(\tau_1-\tau_2)))+i\pi N_g(\zeta_1+\zeta_2)-\frac{\pi^2 T}{4E_c}(2+2\zeta_1\zeta_2)\Bigg].
\eal
Consider now the summation over $\omega_n$. The first term in the summation gives the same form of Eq.~\eqref{eq:expval}, with factor 2. The $\cos(\omega_n(\tau_1-\tau_2))$ part becomes
\bal
\sum_{\omega_n}\frac{-\pi T\nu(1+e^{-|\omega_n|2\ell})\zeta_1\zeta_2\cos(\omega_n(\tau_1-\tau_2))}{|\omega_n|+2E_c\nu/\pi (1+e^{-|\omega_n|2\ell})}\simeq& \sum_{\omega_n}\frac{-\pi T\nu\zeta_1\zeta_2\cos(\omega_n(\tau_1-\tau_2))}{|\omega_n|+2E_c\nu/\pi }\nn\\
\simeq &\sum_{\omega_n}\Big[-\frac{\pi^2 T}{2E_c} +\frac{\pi T\nu|\omega_n|}{(2E_c\nu/\pi)^2}\Big]\zeta_1\zeta_2\cos(\omega_n(\tau_1-\tau_2))\nn\\
=&\frac{\pi^2 T}{2E_c }\zeta_1\zeta_2+\frac{2\pi^2 T^2\nu}{(2E_c\nu/\pi)^2}\zeta_1\zeta_2\Big[\frac{1}{\sin^2(\pi T(\tau_1-\tau_2))}\Big].
\eal
In the first line, we neglect $e^{-|\omega_n|2\ell}$ in the same way that Eq.~\eqref{eq:expval} is approximated by Eq.~\eqref{eq:expval1}. 
Using an approximation
\bal
\exp\Bigg[\zeta_1\zeta_2\frac{2\pi^2 T^2\nu}{(2E_c\nu/\pi)^2}\text{Re}\Big[\frac{1}{\sin^2(\pi T(\tau_1-\tau_2))}\Big]\Bigg]\simeq 1+\zeta_1\zeta_2 \frac{2\pi^2 T^2\nu}{(2E_c\nu/\pi)^2}\Big[\frac{1}{\sin^2(\pi T(\tau_1-\tau_2))}\Big],
\eal
we find that the partition function depends not only $\cos(\pi N_g)$, which is encoded in $\lambda_{{\rm{eff}}}$, but also on $\sin(\pi N_g)$. This yields 
\bal
\frac{Z_2}{Z_0}\simeq &\frac{\lambda_0^2}{2} \left(\frac{2\nu E_c \gamma}{D\pi }\right)^{\nu}\int_0^{1/T} d\tau_1 \int_0^{1/T} d\tau_2 \frac{(\pi T /D)^{2\nu}}{\sin^{2\nu}(\pi T(\tau_1-\tau_2))}\Bigg[4\cos^2(\pi N_g)+4\sin^2(\pi N_g)\frac{2\pi^2 T^2\nu}{(E_c\nu/\pi)^2}\frac{1}{\sin^2(\pi T(\tau_1-\tau_2))}\Bigg]\nn\\
=&\frac{\lambda_0^2}{2} \left(\frac{2\nu E_c \gamma}{D\pi }\right)^{\nu}\frac{1}{T}\int_0^{1/T} d\tau_a \frac{(\pi T /D)^{2\nu}}{\sin^{2\nu}(\pi T(\tau_a))}\Bigg[4\cos^2(\pi N_g)+4\sin^2(\pi N_g)\frac{2\pi^2 T^2\nu}{(E_c\nu/\pi)^2}\frac{1}{\sin^2(\pi T(\tau_a))}\Bigg].
\eal
In the second line, we denoted $\tau_a=\tau_1-\tau_2$, $\tau_b=\tau_1+\tau_2$, where $\tau_b$ is integrated over. 
The first term in the square bracket gives the perturbation, and the second term is the correction corresponds to the charge degeneracy. 

The correction to the entropy is thus
\bal
 S=\frac{\partial}{\partial T}\Big[-T\log\Big(Z_0\Big(1+Z_2/Z_0\Big)\Big)\Big]\simeq \partial_T[-T\log(Z_0)]+\partial_T(-T Z_2/Z_0).
\eal
Including the correction due to the filling factor $\nu$ at $N_g=1/2$, we obtain
\bal
S_{QD}\propto\Big(\frac{T}{E_c}\Big)^2\Big(\frac{T_B}{T}\Big)^{2-2\nu}.
\eal